\documentclass[fleqn,usenatbib]{mnras}

\usepackage{mathptmx}

\usepackage[T1]{fontenc}
\usepackage{ae,aecompl}

\usepackage{graphicx}	
\usepackage{amsmath}	
\usepackage{amssymb}	
\usepackage{lscape} 
\usepackage{url} 

\newcommand{\HI}{H\,{\sc{i}}} 
\newcommand{\MHI}{$\mathrm{M_{HI}}$}
\newcommand{\Msun}{$\mathrm{M_{\odot}}$} 
\newcommand{\kms}{km\,s$^{-1}$}
\newcommand{\Mstellar}{$\mathrm{M_{\ast}}$}

\newcommand{\lyalpha}{Ly$\alpha$}

\title[Continuum and spectral line commensality]{Optimising commensality
  of radio continuum and spectral line observations in the era of the SKA}

\author[Maddox, Jarvis \& Oosterloo]{
Natasha Maddox$^{1}$\thanks{E-mail: maddox@astron.nl},
M.~J. Jarvis$^{2,3}$, T.~A. Oosterloo$^{1,4}$
\\
$^{1}$ASTRON, the Netherlands Institute for Radio Astronomy, Postbus 2,
7990 AA, Dwingeloo, The Netherlands\\
$^{2}$Oxford Astrophysics, Denys Wilkinson Building,
University of Oxford, Keble Rd, Oxford, OX1 3RH, UK\\
$^{3}$Physics Department, University of the Western Cape,
Cape Town, 7535, Republic of South Africa\\
$^{4}$Kapteyn Astronomical Institute, University of Groningen, PO Box
 800, 9700 AV Groningen, The Netherlands\\
}

\date{Accepted XXX. Received YYY; in original form ZZZ}

\pubyear{2016}

\begin{document}
\label{firstpage}
\pagerange{\pageref{firstpage}--\pageref{lastpage}}
\maketitle

\begin{abstract}

The substantial decrease in star formation density from $z=1$ to the
present day is curious given the relatively constant neutral gas
density over the same epoch. Future radio astronomy facilities,
including the SKA and pathfinder telescopes, will provide pioneering
measures of both the gas content of galaxies and star formation
activity over cosmological timescales. Here we investigate the
commensalities between neutral atomic gas (\HI) and radio continuum
observations, as well as the complementarity of the data products. We
start with the proposed \HI\ and continuum surveys to be undertaken 
with the SKA precursor telescope MeerKAT, and building on this, explore optimal
combinations of survey area coverage and depth of proposed \HI\ and
continuum surveys to be undertaken with the SKA1-MID instrument.
Intelligent adjustment of these observational parameters results
in a tiered strategy that minimises observation time while maximising
the value of the dataset, both for \HI\ and continuum science goals. 
We also find great complementarity between the \HI\ and continuum
datasets, with the spectral line \HI\ data providing 
redshift measurements for gas-rich, star-forming galaxies with stellar
masses \Mstellar\ $\sim 10^9$ \Msun\ to $z\sim 0.3$, a factor of three
lower in stellar mass than would be feasible to reach with 
large optical spectroscopic campaigns. 

\end{abstract}



\section{Introduction}\label{sec:Introduction}

The dramatic decrease in the cosmic star formation density 
from $z=1$ to the present day has been well documented
(see \citealt{Hopkins2006}, for example). Conversely, the
neutral gas content (\HI) of galaxies spanning the same redshift range is
much less well known. At $z\sim 0$, large area surveys such as
the Arecibo Legacy Fast ALFA survey (ALFALFA,
\citealt{Giovanelli2005}) and the \HI\ Parkes All-Sky Survey (HIPASS,
\citealt{Barnes2001}, \citealt{Meyer2004}) provide direct detections
of the cold neutral gas in galaxies, from which the \HI\ density of the local
Universe can be derived. Targeted studies, such as the Blind
Ultra-Deep \HI\ Environmental Survey (BUDHIES, \citealt{Jaffe2013}),
or the HIGHz survey \citep{Catinella2015}, have used many hundreds
of hours of observing time to detect several galaxies in \HI\ at $z\sim
0.2$, but the detected objects are not representative of the general
galaxy population, being either located in high density environments,
or lying at the upper end of the galaxy mass distribution. 
An average measurement of the \HI\ content of galaxies not
individually directly detected can be derived from spectral
stacking, which has provided estimates of the cosmic \HI\
density to $z\sim 0.2$ (\citealt{Lah2007}, and \citealt{Rhee2013}, for example), and can be
extended to even higher redshifts.  

At high ($z>1$) redshifts, the neutral gas
content of galaxies and their environments is detected in absorption
along the line-of-sight toward background quasars, resulting in damped
\lyalpha\ systems (DLAs), lower column density Lyman limit systems
and the \lyalpha\ forest. The large numbers of these systems currently
known (several thousands, see \citealt{Noterdaeme2012}) gives an
indirect measurement of the gas content of galaxies at redshifts beyond the reach
of current radio telescopes. The lower redshift limit of $z\sim 1$ is set by the
\lyalpha\ absorption appearing in the ultra-violet, necessitating
expensive space-based spectroscopy.

As \HI\ provides the fuel for star formation, a connection between
star formation activity and the neutral gas reservoirs available to
feed the activity is expected. An intriguing detail is that while the
level of star formation has dropped by more than an order of magnitude
from $z=1$ to the local Universe, the difference between the \HI\
density at $z=0$ measured from \HI\ in emission to $z>1$ measured from
\HI\ in absorption has only changed by a factor of a few (see
\citealt{Rhee2013}, for example). The nearly constant \HI\ density
since $z=1$ indicates replenishment of the gas reservoir, as the gas
consumption timescales are relatively short in comparison. This
discrepancy also points to at least one 
additional ingredient required to connect the reservoir of fuel to
stellar buildup. Molecular gas, as the intermediary between neutral
gas and star formation, is an obvious line of investigation. ALMA,
with its exceptional sensitivity, is revolutionizing the study of
molecular gas, both at low and high redshift. 

Observing \HI\ in emission spanning the redshift range $0<z<1$ 
is within the capabilities of the next generation of radio
telescopes, and is one of the key science cases for the Square
Kilometre Array (SKA, \citealt{Braun2015}). In the meantime, existing
facilities, including the recently upgraded Karl 
G. Jansky Very Large Array (VLA), and the APERture Tile in Focus (APERTIF)
focal-plane array system planned for deployment on the Westerbork
Synthesis Radio Telescope in 2016 \citep{Oosterloo2009}, will allow
measurement of the \HI\ content of galaxies to $z=0.45$ and $z=0.25$,
respectively. A deep \HI\ survey, the COSMOS \HI\ Large Extragalactic
Survey (CHILES) is indeed already underway with the VLA (see Section
\ref{sec:chiles}). 

In preparation for the SKA, mid-frequency pathfinder radio telescopes currently under
construction including South Africa's MeerKAT (\citealt{Jonas2009}) and the
Australian Square Kilometre Array Pathfinder (ASKAP,
\citealt{Johnston2008}), will improve on the capabilities of existing
facilities. Thus, we will be able to begin exploring the gap in
redshift, $0<z<1$, corresponding to lookback times up to nearly 8 Gyr,
which has thus far remained almost entirely inaccessible with respect
to the neutral gas content of galaxies. 

\subsection{Commensality and complementarity}

In practice, the amount of time available on a given facility 
is finite. Performing observations \textit{commensally}, i.e. multiple independent
projects with distinct science goals all using the data from a single
observation, greatly improves the efficiency of a telescope. Clever
adjustment of the observational parameters, including integration
time, survey area, and frequency resolution, enables maximum
scientific return for a given dataset. We focus here on the
commensality between \HI\ and continuum 
observations, with the aim of optimising the scientific return for the \HI\
surveys, by adjusting only the depths and areas of the observations. The
ability to combine surveys with historically different observation
modes, namely spectral line and radio continuum, has only
recently become feasible due to expanded bandwidth of receivers
and sufficient computing to handle the increased data
rates. Previously, observing in spectral line mode was
necessarily restricted to a limited frequency range, corresponding
to a very narrow volume slice, appropriate for targeted
observations. Performing purely commensal continuum observations in
spectral line mode was not worth the increased computing overhead,
because the probability of the small volume probed containing an
\HI-rich galaxy was low. With the increased bandwidth available for
spectral line observations, spanning many hundreds of MHz, the
volume probed for \HI\ detections is vastly increased.

We note that the different computing requirements of spectral line and
continuum science, with respect to imaging, spatial and spectral 
resolution, among other things, will need to be
considered. Spatial and spectral resolution can be customized
for each science case by implementing this flexibility within the
data processing pipeline. Fortunately, the 
\HI\ and continuum observations have similar initial calibration 
requirements, so the computing overhead for processing a set of
observations to be suitable for both \HI\ and continuum science is
minimal. 

Satisfying the observational requirements of multiple science cases
while staying within the boundary conditions of available observing
time requires careful planning
and preparation. We should also be mindful of existing and planned
ancillary data efforts. Survey area, depth, and particularly relevant
for radio observations, spatial resolution, need to be considered. At
the depths probed by next generation radio facilities, the continuum source
density is high \citep{Jarvis2015}, requiring adequate spatial resolution, of the order
of an arcsecond or less, to avoid source confusion. The \HI\ source density
is much lower, and thus confusion is not as problematic.

We also investigate the \textit{complementarity} of the resulting data
products. Ultra-deep neutral hydrogen surveys are planned with both 
MeerKAT (LADUMA, \citealt{Holwerda2012}) and ASKAP
(DINGO\footnote{\url{http://www.physics.uwa.edu.au/~mmeyer/dingo}}),
spending several thousands of hours of observations of single fields to detect
\HI\ in galaxies to higher redshifts and lower \HI\ masses than
possible with existing telescopes. These same observations will be used by the
radio continuum teams, who intend to create ultra-deep
continuum maps using multi-frequency synthesis imaging. The converse
is also true, with \HI\ spectral line data to be extracted from large
area continuum observing programmes. Thus, from the
same observations, simultaneous \HI\ spectral 
line and broad-band continuum data can be obtained, probing both the
neutral gas reservoirs and star formation activity for a large sample
of galaxies spanning a cosmologically significant range in
redshift. This dual-usage of a single dataset to explore two related,
yet distinct, physical processes is unique to upcoming radio surveys.

We focus only on surveys using the MeerKAT pathfinder and the SKA
Phase 1 mid-frequency (SKA1-MID) instrument. We have not included
proposed APERTIF or ASKAP surveys, as the final performance of the phased array
feeds emerging technology has yet to be conclusively measured. Also,
while the single pointing field of view is $\sim$30 deg$^2$, the
frequency range, and thus redshift range accessible for \HI, will not
be as extensive as that of MeerKAT or SKA1-MID.


The current work has two goals. The first is to investigate
commensal observations, focusing on a combination of surveys soon to
begin, in Section~\ref{sec:cases}. We 
then use what we learned from that example to suggest optimal
commensality for upcoming SKA surveys in Section~\ref{sec:SKAdesign},
and focus on parametrization of the \HI\ mass function as a general
science goal. The investigation here serves as input for the ongoing discussion
regarding the definition of the surveys to be undertaken with next generation facilities.
In the second half of the work, starting in
Section~\ref{sec:deepsurveys}, we focus on how \HI\ spectral line and
radio continuum observations are complementary, and what information each provides
with respect to the other. A case study for a pair of surveys
currently underway is in Section~\ref{sec:chiles}. A summary is in
Section~\ref{sec:summary}. Concordance 
cosmology with $H_{0} = 70$ km s$^{-1}$ Mpc$^{-1}$ (thus $h\equiv
H_{0}$/[100 km s$^{-1}$ Mpc$^{-1}$]$=0.7$), $\Omega_{m} = 0.3$,
$\Omega_{\Lambda} = 0.7$ is assumed when computing masses and
luminosities, unless noted otherwise.  

\section{Commensal \HI\ and Continuum Surveys}\label{sec:cases}

Advances in computing power and engineering technology provide
flexibility in observing techniques. In the past,
observations were performed either in spectral line mode, with fine
spectral resolution over a limited frequency range, or continuum mode,
with coarser spectral resolution over a wider frequency
range. Upcoming facilities will have no such restriction, enabling
simultaneous spectral line and continuum observations. While this is
clearly efficient with respect to observing time, the science return
of the observations can also be increased.

\subsection{LADUMA and MIGHTEE with MeerKAT}\label{sec:Laduma}
 
The South African SKA precursor facility, MeerKAT, is a 64-element
array of 13.5\, m dishes located in the Karoo region of South
Africa. The smaller dish diameter of MeerKAT, with
respect to the VLA, results in a larger field of view (FoV), with the primary beam
full width half maximum (FWHM) of nearly 1 degree at 1420MHz for the
former compared to 0.5 degree for the latter. The majority of
operational time has been allocated to several large programmes, with
the remaining time available for smaller individual projects.

Looking At the Distant Universe with the MeerKAT Array (LADUMA), is
the deep \HI\ survey to be undertaken with MeerKAT. LADUMA has been allocated
5000 hours of observing time for a single pointing, to be split
between two frequency bands, covering $0<z<0.58$ for \HI\ (1670--900 MHz) in
Phase 1 operations, and $0.40<z<1.4$ (1015--580 MHz) in Phase 2. 

The MeerKAT International Giga-Hertz Tiered Extragalactic Exploration
(MIGHTEE; \citealt{Jarvis2012}) survey is the MeerKAT continuum key
science survey, and follows a wedding-cake design. The exact
parameters are yet to be finalised, but for our purposes here, we
assume a medium-deep
layer covering 35 deg$^2$, and a deep $\sim$1 deg$^2$ component
coincident with the LADUMA pointing. The final area coverage will not
change significantly from these values. The observations allocated to LADUMA will be
made available to the MIGHTEE team for continuum science, while the
data collected for MIGHTEE can be used for \HI\ science, as it will
be collected in spectral line mode, with velocity resolution of
26kHz, corresponding to 5.5 \kms\ at $z=0$ for \HI, increasing
  to 7.8 \kms\ at $z=0.4$. Thus, LADUMA and MIGHTEE are to be observed
  entirely commensally.

\subsubsection{LADUMA as MIGHTEE Deep}

The 5000 hours of integration for LADUMA is sufficient to directly
detect the most \HI-massive systems at $z=0.6$ and beyond.
This does not, however, directly imply that the continuum data will
reach depths corresponding to such long integration times. The longest
planned baseline of the MeerKAT array is 8\,km, resulting in spatial resolution 
ranging from $\sim$\,6\arcsec\ at 1420MHz to $>$15\arcsec\ at
600MHz. Using the approximation listed in equation 27 in
\citet{Condon2012} to compute the noiseless confusion distribution
width $\sigma^{\star}_c$, the continuum confusion noise is
$\sim$1$\mu$Jy beam$^{-1}$ for these resolutions, which is reached in
approximately 50 hours with MeerKAT, and cannot be overcome without introducing
longer baselines.

A more significant contribution that LADUMA data can make
to the MIGHTEE project is by providing redshifts for the \HI-rich,
but relatively low stellar mass galaxies at $z<0.4$, which will
have significant levels of star formation, but are optically faint,
and therefore difficult spectroscopic targets at visible wavelengths.

\subsubsection{MIGHTEE as LADUMA Wide}

While LADUMA provides a deep continuum pointing for MIGHTEE, MIGHTEE
provides complementary wide-area \HI\ data with observations 
covering 35 deg$^2$, with tens of hours of integration per
pointing. This is clearly not comparable to the thousands
of hours devoted to LADUMA, but it is sufficient to
detect the most massive \HI\ systems, which will be under-represented
in the relatively small volume contained within a single
pointing. These same gas-rich galaxies will also have significant star
formation, and thus will be a prominent population within the MIGHTEE
continuum dataset. We will then have simultaneous measures of the
neutral gas content and star formation rates of galaxies extending
beyond $z\sim 0$, including a wide range of environments, as the
volume probed is large enough to include a number of massive clusters.
The benefits of using the medium-deep continuum survey
data to supplement \HI\ surveys is expanded and investigated in
the following section. 

\subsection{Galaxy counts and redshift distribution}\label{subsec:galcounts}

Here we make qualitative predictions
of what we can expect from planned surveys and how we can best combine
the available information. In the absence of direct observations of
the \HI\ content of galaxies at $z>0.2$, we rely on information
gathered from the low redshift Universe, combined with simulations, to estimate the
number of galaxies, and their distribution in mass and redshift,
upcoming surveys will detect.

We start by investigating the LADUMA--MIGHTEE survey combination,
focusing on the \HI\ aspect, as this is where the greatest
complementarity is seen. The confusion issues for the continuum
observations make a similar analysis much more complex. For the
following, we use the information provided to the public for 
MeerKAT\footnote{\url{http://public.ska.ac.za/meerkat/meerkat-schedule},
  \\ Released 08 Dec 2015}. The listed System Equivalent Flux Density
(SEFD) values indicate MeerKAT's performance is expected to be better
than that of the VLA, possibly by as much as a factor of two in sensitivity.

As listed in Table~\ref{tab:SKA1}, for the low redshift (Phase 1)
component of LADUMA, the integration 
time will be 1000 hours, which corresponds to a 5-$\sigma$\ \HI\ flux
limit of 0.015 Jy \kms\ for MeerKAT over 250 \kms. We choose 
this profile width as we expect the majority of detected objects to be massive systems,
with profiles that can easily span several hundreds of \kms.  Equivalently, for 24 hours per
MIGHTEE pointing, the 5-$\sigma$\ \HI\ flux limit is 0.07 Jy \kms. We assume that
the full 35 deg$^2$ area will be composed of 35 individual pointings,
instead of a single mosaic, which removes the complication of an increasing 
effective exposure time for a given patch of sky within a
mosaic, as the primary beam increases in size going to lower
frequency, and hence to higher redshift, and the individual pointings increasingly overlap. We
note, however, that in practice there will be overlap between some of
the pointings, thus the calculations here are conservative with
respect to the mass detection limit.

We use the parameters from the analytic fits to the $dN/dz$ relations
derived from the simulations of \citet{Obreschkow2009} to convert 
flux limits in \HI\ to numbers of galaxies detected as a function of redshift. The
simulations are consistent with the observed $z=0$ \HI\ mass
function and CO(1--0) luminosity function, and allow for evolution of
the \HI\ content of galaxies with redshift.  The $dN/dz$ relations
enable us to determine, for a given redshift shell and sky area, the
total number of galaxies expected to be detected above a specified flux limit. Appropriate
relations for arbitrary integrated flux limits can be obtained
by interpolating the function coefficients for the given set of flux
limits. We note that the $dN/dz$ relations are based on the 
cosmological parameters of  $H_{0} = 100h$ km 
s$^{-1}$ Mpc$^{-1}$ with $h=0.73$,  $\Omega_{m} = 0.25$ and
$\Omega_{\Lambda} = 0.75$ (\citealt{Spergel2003}), which have been
superceded by, for example, \citet{Komatsu2011}, but the effect on the
results is negligible. 

\begin{figure}
\includegraphics[width=\columnwidth]{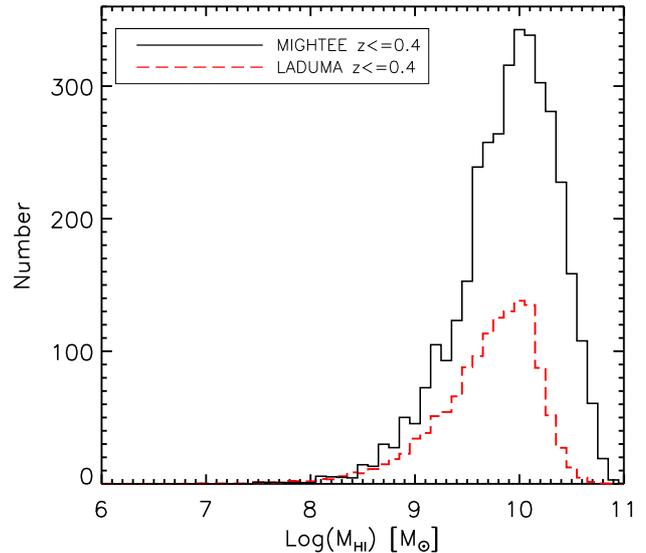}
\caption{Number of galaxies as a function of \HI\ mass observed in the
  single pointing of LADUMA, (red dashed line), and in the 35 deg$^2$ of
  MIGHTEE (black solid line), both for $0<z<0.4$.} 
\label{fig:Lad_Mig_zhist}
\end{figure}

We also make use of the \HI\ mass function (HIMF) derived at $z\sim 0$ from
ALFALFA \citep{Martin2010}, which spans $6<$Log(\MHI)$<11$. Galaxies
with \MHI$>10^{11}$\,\Msun\ are exeedingly rare, so we do not include
them in our calculations. From the \HI\ mass function, 
we can estimate how the galaxies in each redshift shell are
distributed in mass. For the single MeerKAT pointing of LADUMA 
and 35 deg$^2$ of MIGHTEE, we can compute how many
galaxies of each mass we expect to observe over the full survey
redshift range. Fig.~\ref{fig:Lad_Mig_zhist} shows the result of this
exercise, limited to $z<0.4$. As seen, the continuum survey, MIGHTEE, 
detects nearly three times as many galaxies in \HI\ as the deep \HI\ survey,
LADUMA, due to the larger area, and thus larger volume, surveyed. The
difference is particularly notable at the high \HI-mass end, where
very few massive \HI\ systems are detected by LADUMA. 

\begin{figure}
\includegraphics[width=\columnwidth]{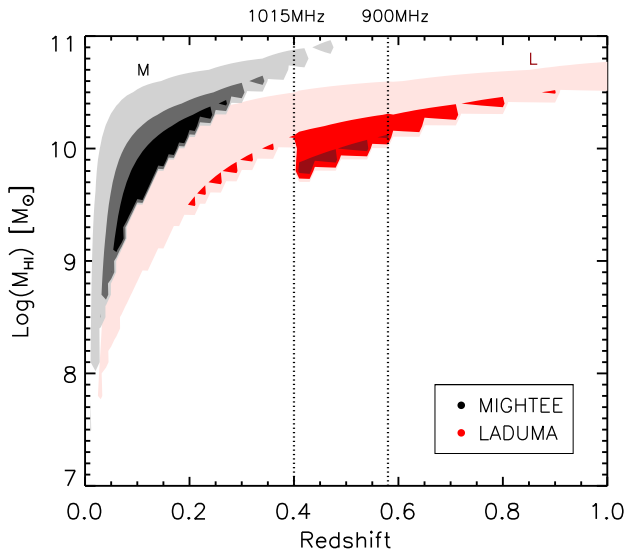}
\caption{The coverage in the redshift--\MHI\ plane for LADUMA (red
  contours, with `L' label) and MIGHTEE (greyscale contours, with `M'
  label). The parameter space is 
  divided into cells of width 0.01 in redshift and 0.1 in
  Log(\MHI). The contour levels for both surveys are the same, 
  surrounding regions of parameter space containing at least
  one (faint shading), 10 (medium shading), and 25 (dark shading)
  galaxies per cell. Due to the overlapping frequency coverage of the two 
  MeerKAT receivers, the redshift range $0.40<z<0.58$ is observed for
  the full 5000 hours for LADUMA, and is correspondingly deeper.}
\label{fig:Lad_Mig_MHI_z}
\end{figure}

Fig.~\ref{fig:Lad_Mig_MHI_z} shows how the detected galaxies counted
in Fig.~\ref{fig:Lad_Mig_zhist} are distributed in the redshift--\MHI\ plane. 
The plane is divided into cells of width $\Delta z$=0.01 and
$\Delta$Log(\MHI)=0.1. The contours for both LADUMA and MIGHTEE are
set at 1, 10 and 25 galaxies per cell. Complications due to spatial
resolution determined by the size of the synthesized beam, how 
this size changes with redshift, and the resulting effect on column density
sensitivity have been neglected for this exercise. The sensitivity is
maximized when the size of the galaxy matches the size of the beam,
and decreases for larger and smaller sources. Therefore, the lower
boundaries of detected galaxies are specifically for galaxies the same
size as the beam, and extends to higher masses for galaxies of smaller
or larger physical sizes.

The two MeerKAT surveys are highly complementary, with MIGHTEE primarily detecting
high \HI\ mass galaxies at low redshift, while LADUMA reaches further
down the \HI\ mass function. This access to wide
area will constrain the high mass end of the \HI\ mass function, which
is populated by galaxies with low space density. The wide area is
also essential for studies of galaxy environment and large scale
structure, which are inaccessible to the narrow, deep observations. 
Thus, including \HI\ data from the continuum survey greatly enhances
the potential scientific output of the \HI\ studies. 

\section{SKA Phase I Commensal Survey Design}\label{sec:SKAdesign}

A tiered, or ``wedding-cake", survey design, is often employed by large
projects, as it provides dynamic range in depth, while also probing a
large volume. The relative depths of the tiers must be set to
provide optimal coverage of some parameter of interest. Here we
investigate surveys suggested to be undertaken with 
SKA1-MID. The facility will be a combination of 133 15\,m new
  SKA dishes, in addition to the 64 13.5\,m MeerKAT dishes, as outlined
in the Baseline Design Document Version
2\footnote{\url{https://www.skatelescope.org/key-documents/}, \\
  Released October 2015}. As the final telescope will be composed of two types of dishes, each
with their own performance characteristics, accurately anticipating
the sensitivity of SKA1-MID is difficult. However, given the SEFD
values provided in the Baseline Design 
Document for the SKA dishes, and those provided in the public MeerKAT
documentation, we conservatively estimate a factor of 5 increase
in sensitivity of SKA1-MID Bands 1 and 2, with
respect to the VLA, consistent with simulations by \citet{Popping2015}.

We construct the survey tiers keeping in mind that the
resulting data will be used commensally by both the \HI\ and continuum teams, and
thus restrict ourselves to the range of parameters that are useful to
both. We use the sensitivity to \HI\ mass as a function of redshift as a measure of
the value of the survey tier combination, and optimise the tiers by
rearranging their depth and area to maximise coverage of the
redshift--\MHI\ plane. 

\subsection{Optimal SKA Survey Tiers}\label{subsec:commensal}

The redshift and \HI\ mass distribution of galaxies expected from
surveys displayed as in Fig.~\ref{fig:Lad_Mig_MHI_z} can be used as a
powerful diagnostic tool to explore combinations of depth and area
which result in the most scientifically useful dataset. LADUMA and
MIGHTEE are clearly well paired at $z<0.4$, enhancing the usefulness
of the data.

The SKA \HI\ science working group (SWG) proposed a tiered observing strategy to
be undertaken with SKA1-MID to address many extragalactic \HI\ science cases,
consisting of an ultra-deep, small area survey, a medium-wide,
medium-deep survey, and a wide-area, relatively 
shallow survey. We refer to these three tiers as Ultra-Deep (UD),
Medium-Wide (MW) and Wide (W). Their areas, integration times and
approximate 5-$\sigma$\ sensitivities that can be expected are listed
in Table~\ref{tab:SKA1} as the ``Fiducial'' case, as set by \citep{Lister2015}. 
A more detailed discussion within the \HI\ SWG of the 
three tiers will come at a later date, when the technical specifications of the
SKA1-MID are better known, but the suggested Fiducial case serves as a useful
starting point. We have not included the ``strawman''
surveys outlined in the SKA \HI\ galaxy evolution chapter
(\citealt{Blyth2015}), as they rely heavily on the 
SKA-SURVEY instrument, construction of which has been deferred.

\begin{figure}
\includegraphics[width=\columnwidth]{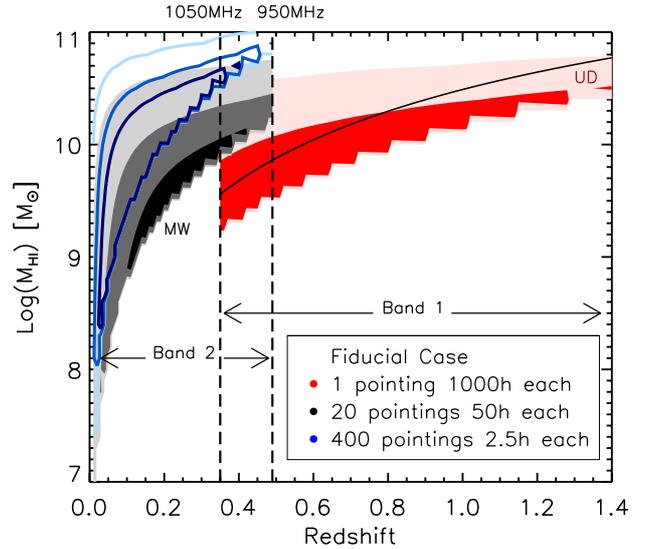}
\caption{The coverage in the redshift--\MHI\ plane for the Fiducial
  SKA1 UD (red filled contours, with `UD' label), MW (greyscale filled
  contours, with `MW' label), and W
  survey (blue open contours). The parameter space 
  is the same as in Fig.~\ref{fig:Lad_Mig_MHI_z}, but extends to $z=1.4$. The contour levels
  for all three surveys are the same, surrounding regions of parameter
  space containing at least 1 (faint shading), 25 (medium shading),
  and 100 (dark shading) galaxies per cell. The contours for the wide
  survey are unfilled to show the overlapping regions between W and MW. 
  The vertical dashed lines indicate the lower edge of Band
  1 at 1050MHz, and the upper edge of Band 2 at 950MHz. The black solid curve
  indicates the \HI\ mass sensitivity if the performance of Band 1 is
  reduced by a factor of two.}
\label{fig:SKA1_proposed1_MHI_z}
\end{figure}

We show in Fig.~\ref{fig:SKA1_proposed1_MHI_z} the coverage in the
redshift--\MHI\ plane resulting from the three Fiducial \HI\ survey
tiers. Increasing the integration
time to 4000 hours per tier, to be consistent with the proposed surveys
discussed below, improves the sensitivity by a factor of two, but the
general features remain unchanged.

A critical aspect missing from the \HI\ SWG observing
strategy is the possibility of observing commensally, in particular
sharing observations with the SKA continuum SWG
(\citealt{Prandoni2015}). While commensal observing
necessarily involves some compromises to be made on some aspects,
including field selection, the benefits are significant, including the
increased available observing time.

The continuum SWG also proposed a three-tiered
observing strategy, with areas comparable to those
explored here, with a single pointing UD tier, a MW tier spanning
10--30 deg$^2$, and a W tier of $\sim$\,1000 deg$^2$. The suggested
continuum RMS values for the three tiers of 0.05, 0.2 and 1$\mu$Jy,
respectively, can be compared with the estimated RMS values computed
for the Fiducial and Optimised surveys as listed in
Table~\ref{tab:SKA1}. We do not list continuum RMS values for the
Requested or Ideal cases, because the relevant receivers do not
currently exist.

We suggest a number of tiered survey strategies working
within the framework of commensal extragalactic \HI\ and continuum
observations. The three tiers under consideration are driven by
different science cases. The UD tier is driven primarily by the redshift evolution of
\HI\ science case, and has very little flexibility. In order to 
detect the faintest objects at the greatest distances, long
integration times are required. The deepest integration is achieved by
observing only a single pointing, the area of which is set by the FoV
of the telescope. At the other extreme, the W tier is of general use for many science
cases. The size is restricted by the amount of area that can be
covered with individual pointings long enough such that they are not dominated by
calibration overheads. The MW tier is 
intermediate between the two, and requires advance planning to
maximise the scientific return. The area and depth must be
simultaneously adjusted such that a representative volume containing
detectable galaxies is observed.

For our purposes, we assume in the following that the continuum
science will be the primary driver behind the MW tier,  
as it will provide the bulk of the continuum science return. 
The W tier is too shallow to explore a signficant redshift range,
while the narrow UD tier will suffer from cosmic variance, which becomes
a significant source of uncertainty for areas smaller than about 10
deg$^2$ \citep{Heywood2013}. This is consistent with the SKA continuum SWG
specification that the MW tier should be 10--30 deg$^2$, while the 
required sensitivity should be sufficient to sample the peak of star
formation, at $1<z<3$.

Continuum confusion is not an issue for SKA1-MID as it is 
for MeerKAT and ASKAP, as the array layout will enable sub-arcsecond
spatial resolution. \HI\ science can employ a range of spatial
resolutions, matching the resolution with the angular scale of the
targets. We assume that the continuum and \HI\ data processing
pipelines will be able to customize the spatial resolution of the
respective maps for optimal sensitivity.

\begin{table}
\centering
  \caption{Parameters for the two MeerKAT surveys, LADUMA (Phase 1) and MIGHTEE,
    and the four different combinations of
    survey tiers suggested for the SKA Phase 1 MID instrument. The
    confusion noise floor of $\sim$1$\mu$Jy for MeerKAT limits the continuum depth
    of LADUMA and MIGHTEE. The Fiducial case is
  from the SKA \HI\ Science Working Group, whereas the Optimised case
  alters the depth and areas of these three tiers. The Requested case
  reflects the frequency coverage of the suggested revised Band 1 and
  Band 2, while the Ideal case is for a hypothetical, single-octave
  receiver system. For the Optimized and Ideal cases, 
  the UD tier is two pointings covering the same redshift range, to
  alleviate cosmic variance. For the Requested case, the two UD
  pointings are split between Bands 1 and 2. T$_{exp}$ is the
  effective integration time per pointing, and the \HI\ 
flux limits assume SKA1-MID sensitivity 5 times that of the VLA 
and profile widths of 250\,\kms. Continuum RMS values
for the UD tiers assume Band 1, and Band 2 for MW and Wide. The
combination of lower SEFD and wider available bandwidth for Band 2
with respect to Band 1 results in similar RMS values for the UD and MW
tiers for the Optimised case.}
\label{tab:SKA1}
\begin{tabular}{lcccc} \hline
Name & Area & T$_{exp}$ & \HI\ 5-$\sigma$ & Continuum RMS \\
 & deg$^2$ & hours & Jy \kms & $\mu$Jy \\ \hline
LADUMA & 1 & 1000 & 0.015 & 1 \\
MIGHTEE & 35 & 24 & 0.07 & 1 \\ \hline
Fiducial & & & & \\
UD & 1 & 1000 & 0.003 & 0.07\\
MW & 20 & 50 & 0.014 & 0.14\\
Wide & 400 & 2.5 & 0.06 & 0.64\\ \hline
Optimised & & & & \\
UD & 2 & 2000 & 0.002 & 0.05\\
MW & 10 & 400 & 0.0047 & 0.05\\
Wide & 1000 & 4 & 0.05 & 0.51\\ \hline
Requested & & & & \\
UD & 2 & 2000 & 0.002 & \\
MW & 30 & 50 & 0.014 & \\
Wide & 1000 & 4 & 0.05 & \\ \hline
Ideal & & & & \\
UD & 2 & 2000 & 0.002 & \\
MW & 30 & 50 & 0.014 & \\
Wide & 1000 & 4 & 0.05 & \\ \hline
\end{tabular}
\end{table}

A critical aspect of the following investigation is the frequency
coverage of the SKA1-MID instrument. The current design for the
mid-frequency receivers is for Band 1 to cover 1050--350MHz
(corresponding to $0.35<z<3$ for \HI\ observations), and Band 2
1700--950MHz ($z<0.5$), the extent of which are indicated as vertical
dashed lines in Fig.~\ref{fig:SKA1_proposed1_MHI_z}. \HI\ surveys are
particularly affected by the choice of frequency coverage, as it
directly dictates the redshift range that can be surveyed in a given
observation. This is highlighted by the discontinuity in coverage in
Fig.~\ref{fig:SKA1_proposed1_MHI_z} at 1050MHz, or $z<0.35$, indicating a
mismatch in depth between the UD and MW components. Constructing a
receiver band such as the current planned Band 1 is a 
challenging engineering task. We have thus marked on
Fig.~\ref{fig:SKA1_proposed1_MHI_z} 
the equivalent \HI\ mass sensitivity if the Band 1 performance is
decreased by a factor of two. Current measurements of the receiver
indicate it performs slightly better than this at low redshift (high
frequencies), with decreasing sensitivity to high redshift (lower
frequencies)\footnote{\url{http://astronomers.skatelescope.org/ska1/}}. 

Alternative possibilities for receiver bands are currently under
discussion. One option that has been suggested is to replace the
current wide Band 1 with a single-octave band covering 875--475MHz
($0.6<z<2.0$ for \HI), and the low redshift Band 2 with a second band
covering 1470--795MHz
($0<z<0.8$)\footnote{
  \url{https://skaoffice.atlassian.net/wiki/display/EP/ECP+Register}\\
ECP150027}. This arrangement shifts the
overlapping redshift range from $0.35<z<0.5$ to $0.6<z<0.8$.
This arrangement is listed as ``Requested'' in Table~\ref{tab:SKA1}.

Addtional options involve a single-octave band covering either 1420--710MHz
($0<z<1$ for \HI), or 1200--600MHz ($0.2<z<1.4$). Although the second
range excludes observing \HI\ at $z=0$, it is an attractive option
for a number of reasons. First, with respect to survey commensality,
detailed \HI\ observations of nearby galaxies have different
requirements to the deep \HI\ surveys aiming for higher redshift, so
it is sensible to perform the $z=0$ and $z>0$ surveys separately. In
addition, there is known severe RFI at
1170--1300MHz\footnote{\url{http://public.ska.ac.za/meerkat/meerkat-schedule}},
rendering extracting data from observations within this frequency range very
problematic. Continuum studies also stand to benefit, as the lower
frequency limit of 600MHz results in a larger FoV and enhances
sensitivity due to the spectral shape of radio sources, albeit with poorer spatial
resolution. Including coverage at $1.0<z<1.4$ at the expense of
$z<0.2$ also offers the opportunity to explore new parameter space,
which is only accessible to the SKA. The number of
galaxies directly detected at such high redshifts will be modest, but
we will gain the most new information with such observations. 

In the following, we investigate three cases for frequency coverage. The first
assumes the frequency bands as they are currently defined, but
optimising survey tier depth and area (Optimised). The second
refers to the proposed shifting of the frequency coverage of Bands 1 and 2
(Requested), and the third assumes a single band covering either
1420--710MHz or 1200--600MHz (Ideal). We restrict the total 
observing time for each survey tier to be nominally one year, which corresponds to
approximately 4000 hours of usable integration time. 

For the Optimised case, we first fix the UD component, which has very few options for
variation. We choose to split the 4000 hours into two pointings
instead of one, to alleviate cosmic variance. We are then free to
adjust the area and exposure time per pointing of the MW and W
components to address the science cases. As a
general rule, increasing area improves coverage of the high mass
region of redshift--\MHI\ parameter space, whereas increasing integration
time improves coverage at lower masses.

\begin{figure}
\includegraphics[width=\columnwidth]{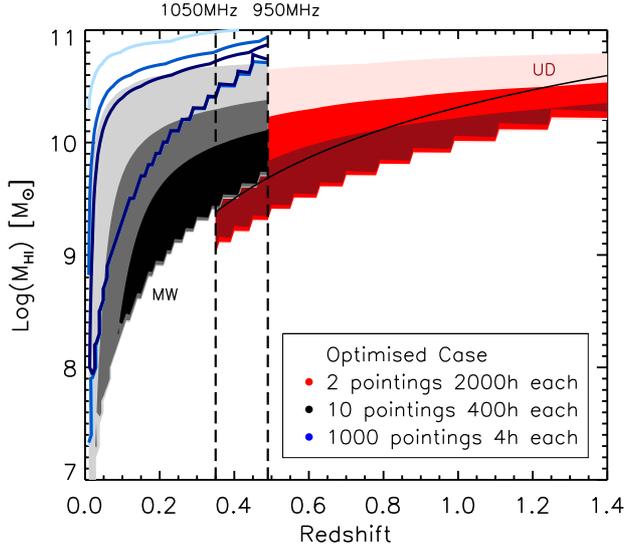}
\caption{The coverage in the redshift--\MHI\ plane for an SKA1 UD
  (red filled contours, with `UD' label), MW (greyscale filled
  contours, with `MW' label), and W survey (blue open contours), corresponding to the
  Optimised case in Table~\ref{tab:SKA1}. The contour levels
  for all three surveys are the same, surrounding regions of parameter
  space containing at least 1 (faint shading), 25 (medium shading),
  and 100 (dark shading) galaxies per cell. The black solid curve
  indicates the \HI\ mass sensitivity of Band 1 if it is reduced by a
  factor of two. The characteristic mass, which is $M_{\star} =
  10^{9.96}$\Msun\ at $z=0$, is well covered with this set of surveys.}
\label{fig:SKA1_best1_MHI_z}
\end{figure}

Fig.~\ref{fig:SKA1_best1_MHI_z} shows how the redshift--\MHI\
plane is populated for the Optimised surveys listed in Table~\ref{tab:SKA1}, after
adjusting the three tiers for optimal coverage in mass and
redshift. For this arrangement, in 
order to maintain high dynamic range in \HI\ mass at low redshifts,
the MW tier is necessarily deep, with correspondingly smaller area to
keep the total survey time within the 4000 hour budget. 10 deg$^2$ is
at the low end of the range of survey areas deemed to be acceptable
for this tier by the SKA continuum SWG (\citealt{Prandoni2015}).

We have increased the area of the wide tier from 400 to 1000 deg$^2$, which
better samples the high-mass end of the \HI\ mass
distribution. Increasing the area further would require shorter
exposure times per pointing, which we want to avoid. As \HI\ column density
sensitivity also scales with exposure time, longer exposures have
greater sensitivity to diffuse \HI, and therefore can be imaged
at higher spatial resolution, than shorter exposures. Short
integration times over a larger area, with the accompanying poorer
resolution, would not be a significant improvement over currently
available technology, and would not efficiently exploit the full
capabilities of the SKA.

\begin{figure}
\includegraphics[width=\columnwidth]{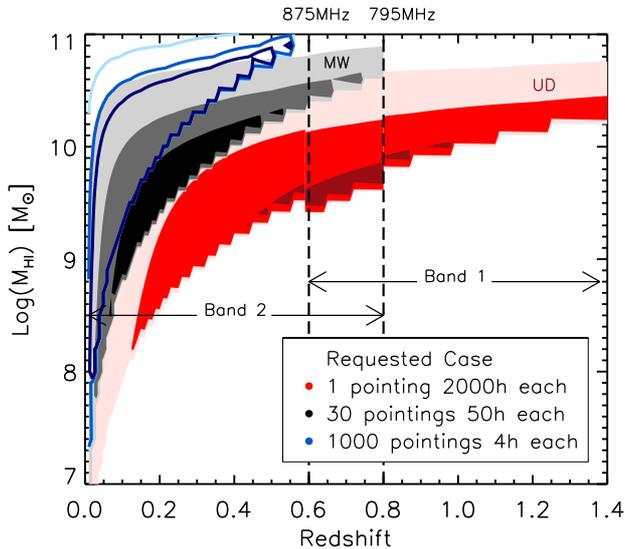}
\caption{Same as in Fig.~\ref{fig:SKA1_best1_MHI_z}, but 
  for the frequency band definition corresponding to the Requested case.
  Note that the UD component now must be built from two separate
  integrations, one with Band 1 and one with Band 2, while the
  redshift range $0.6<z<0.8$ is observed twice.}
\label{fig:SKA1_ECR_MHI_z}
\end{figure}

Fig.~\ref{fig:SKA1_ECR_MHI_z} shows a combination of survey tiers
more appropriate for the Requested frequency coverage. In order to
span the full redshift range, the 
UD tier has been split in two, with one 2000h integration at low
redshift (Band 2), and one 2000h integration at high redshift (Band 1). The redshift
range $0.6<z<0.8$ is covered by both Band 1 and Band 2, and thus
receives 4000h of integration, and is subsequently deeper. The primary difference
between the Optimised and Requested cases is in the MW tier,
which no longer needs to be as deep. In fact, the coverage of the MW tier is achieved by
using only 30 pointings of 50 hours each, using only 1500 of the
nominally available 4000 hours. 30 deg$^2$ is also well-matched to the
planned survey area of the LSST deep fields, which will cover 36
deg$^2$ in the first instance. The remaining time beyond 1500 hours
could be used to more than double the area covered, if desired.

\begin{figure}
\includegraphics[width=\columnwidth]{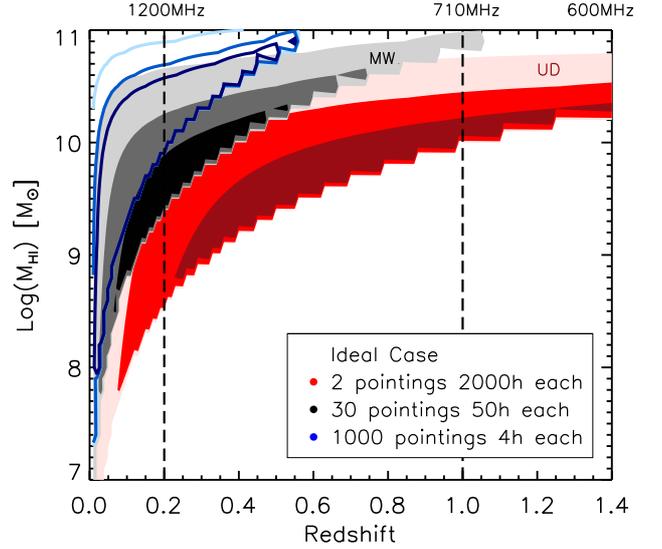}
\caption{Same as in Fig.~\ref{fig:SKA1_best1_MHI_z}, but 
  for the frequency band definition corresponding to the Ideal in
  Table~\ref{tab:SKA1}. Note that either $0<z<1$ or $0.2<z<1.4$ can be
  observed at once, but not the full redshift range $0<z<1.4$. The
  lack of the $z<0.35$ lower redshift cutoff for the UD component means the
  MW tier can be shallower and wider, and completed in only 1500
  hours. The absence of the $z>0.5$ upper redshift cutoff enables the
  MW tier to continue contributing coverage of the high-mass galaxies
  up to $z=1$. Either band definition results in a more
  efficient set of surveys which covers more of the redshift--\MHI\
  plane than with the existing Bands 1 and 2.}
\label{fig:SKA1_best2_MHI_z}
\end{figure}

Fig.~\ref{fig:SKA1_best2_MHI_z} shows the alternative arrangement for
frequency coverage, 
listed as Ideal in Table~\ref{tab:SKA1}, using either the 
1420--710MHz or 1200--600MHz single band. Although we display the full
redshift range of $0<z<1.4$, note that only either the range $0<z<1$
or $0.2<z<1.4$ can be covered with a single observation. 
The MW and W tiers are unchanged from the Requested case, but the UD
tier is again composed of two, 2000h integrations, which span the full
available redshift range.

We emphasize here that, between the Optimised and Requested or Ideal
frequency setups, it is the MW tier that is the most effective for
optimizing the coverage of the redshift--\MHI\ plane. As 
the specifications of the MW tier are of critical importance to the
continuum SWG, the importance of full investigation of
survey commensality becomes apparent.

We have thus far not mentioned the nearby galaxies extragalactic \HI\
science case, which aims to map the neutral gas distribution of a
number of local, well-resolved galaxies in much greater detail than
will be achieved with either the ultra-deep or wide-area
surveys. These targeted observations will observe a small number of
fields with long exposure times, similar to the parameters for the MW
surveys discussed here. However, these MW observations cannot be
easily shared with the local galaxies communities, as the fields would be
specifically chosen such that they do \textit{not} have prominent nearby galaxies
within them. Although the spectral dimension allows us to see
in \HI\ beyond the local galaxies, they are optically opaque,
rendering the entire volume beyond the galaxies inaccessible to
ancillary data. While there is no clear commensality between
  the deep extragalactic surveys discussed 
here and the resolved studies of nearby galaxies, the \HI\ and
continuum teams targeting these galaxies will certainly exploit the
high degree of commensality between their respective observations.

\subsection{Parametrizing the \HI\ Mass Function}

With several options for survey design available to us, we need some
way of quantifying the quality and usefulness of the resulting dataset. 
Population studies require dynamic range in mass at a given redshift,
whereas evolution studies require dynamic range in redshift at a given mass.
\HI\ content as a function of environment requires sufficient survey
volume to include a range of galaxy group and cluster masses. 
Applications such as investigating the evolution of the \HI\ mass
function with redshift would benefit from coverage of the
redshift--\MHI\ plane spanning the break of the relation, which is at
Log(\MHI) = 9.96 \Msun\ at $z\sim 0$ \citep{Martin2010}. If continuous
coverage of \HI\ mass at a given redshift is a priority, then
adjusting exposure times and survey areas can minimize gaps in the
redshift--\MHI\ plane.

We have chosen as an illustrative target science goal the characterisation of the
\HI\ mass function (HIMF) as a function of redshift. This fundamental
relation can be integrated to give the cosmic \HI\ density,
$\Omega_{\mathrm{HI}}$, as a function of redshift. Knowledge of the
HIMF not only provides information about the \HI\ content of galaxies,
but it provides the contextual framework within which other \HI\
studies are conducted, i.e. are the galaxies under investigation
chosen from the rare, high-mass end of the distribution, or are they
relatively common? Quantifying the parametrization of the HIMF at several redshifts
allows us to move beyond the so-called ``strawman'' surveys, and set
realistic guidelines for a combination of survey tiers that results in
a coherent dataset which maximizes scientific return while minimizing
duplicated effort.

We use the Schechter function parametrization of the HIMF as used by
the ALFALFA team in \citet{Martin2010}, given by:

\begin{equation}
\phi(M_{\mathrm{HI}}) = \frac{dn}{d \mathrm{log} M_{\mathrm{HI}}} = \mathrm{ln} 10\,
\phi_{\star} \left(\frac{M_{\mathrm{HI}}}{M_{\star}}\right)^{\alpha +
  1} e^{- \frac{M_{\mathrm{HI}}}{M_{\star}}} \\
\label{eq:HIMF}
\end{equation}

\noindent There are three parameters to determine: the faint-end slope $\alpha$,
the characteristic mass $M_{\star}$, and the normalisation
$\phi_{\star}$. The lack of observational data regarding the detailed distribution
of \HI\ content of individual galaxies at $z>0.2$ provides little
constraint on any evolution in the HIMF, so we do not include any
evolution in the three parameters. The
combination of observing \HI\ in emission at low redshift, and high redshift
observations of damped \lyalpha\ absorption seen in the spectra of
background quasars constrains the evolution of $\Omega_{\mathrm{HI}}$
to be at most a factor of a few between $0<z<2$, so we do not expect
drastic evolution of the individual HIMF parameters over our redshift range of interest,
$0<z<1.4$. Simulations from \citet{Dave2013} indicate that the high mass
end of their HIMF steepens significantly, but only at high redshifts, $z>2$.

\begin{figure}
\includegraphics[width=\columnwidth]{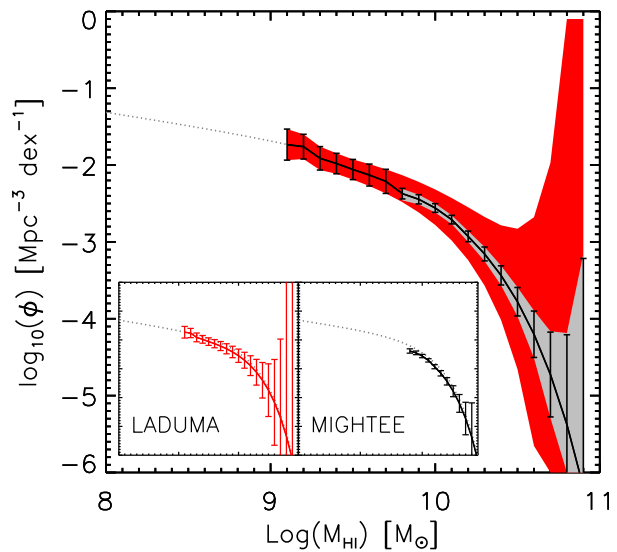}
\caption{The HIMF resulting from combining \HI\ data from MIGHTEE and 
  LADUMA at $0.1<z<0.2$. The two insets show the HIMF from MIGHTEE
  and LADUMA individually, and the main panel shows is the combination of the
  two. The x-axis in the main panel and the insets span the same
  range. The grey region is the HIMF from MIGHTEE 
  including uncertainties, the red region is from LADUMA with
  the associated uncertainties, and the black line with errorbars is the combination of the
  two. The dotted grey line here and in the following figures is the
  input $z=0$ HIMF from ALFALFA, for 
reference.}
\label{fig:Lad_Mig_HIMF_inset}
\end{figure}

We construct HIMFs from the combinations of surveys introduced in
Section~\ref{subsec:commensal} and see how the different surveys
provide complementary constraints. As an illustration, the advantage
gained from combining surveys of different 
depth and area is displayed in Fig.~\ref{fig:Lad_Mig_HIMF_inset},
which shows the HIMF resulting from the
MIGHTEE  and LADUMA surveys individually, and the result from
combining the data. The uncertainties on the individual HIMFs from
LADUMA and MIGHTEE are $1/\sqrt{N}$, where $N$ is the number of
objects in the given mass bin of width $\Delta$Log(M) = 0.1. These are
represented by the shaded areas in the main panel, and the final 
combined HIMF is shown in black with errorbars. The large volume covered by
MIGHTEE is entirely responsible for constraining the high-mass end of
the HIMF, as LADUMA does not contain any of these rare
galaxies. At the low-mass end, the deep flux limit of LADUMA is
required, and a small mass range around the break in the relation is
covered by both surveys. When combined, the
three parameters describing the HIMF are better constrained than
considering each survey individually.

\begin{figure}
\includegraphics[width=\columnwidth]{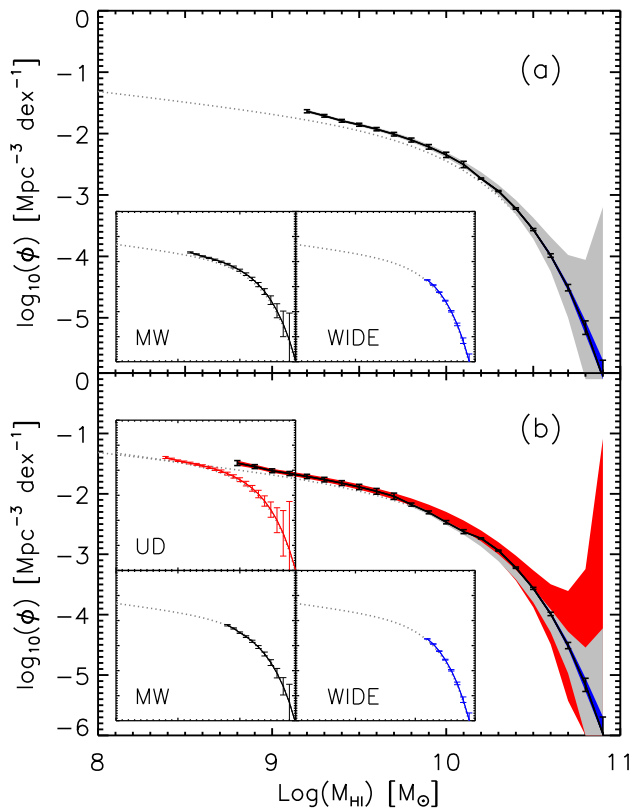}
\caption{The HIMF resulting from combining \HI\ data from the MW (grey) and W (blue)
components of the Optimised survey (top panel), and the UD (red), MW (grey) and W (blue)
components of the Ideal surveys (bottom panel), both at
$0.2<z<0.3$. The UD component is not included in the top panel as Band
1 does not extend to these low redshifts. The addition of the UD
component in the bottom panel shows the HIMF can be probed to lower
masses. The x-axis in the main panel and the insets span the same
  range.}
\label{fig:best1_best2_HIMF_z0203_inset}
\end{figure}

Fig.~\ref{fig:best1_best2_HIMF_z0203_inset} 
compares the HIMF that can be constructed from the Optimised surveys
and either of the Ideal options, at $0.2<z<0.3$. For
the top panel, the existing Band 1 does not cover
redshifts below $z<0.35$, so the UD component is excluded. The break
and the high-mass end of the HIMF are well constrained by the MW and W
survey components, as seen by the small uncertainties. Alternatively, 
the Ideal cases in the bottom panel do incorporate the UD component,
which has large uncertainties at high \HI\ masses, but extends the
HIMF to lower masses, helping constrain the low-mass slope.

\begin{figure}
\includegraphics[width=\columnwidth]{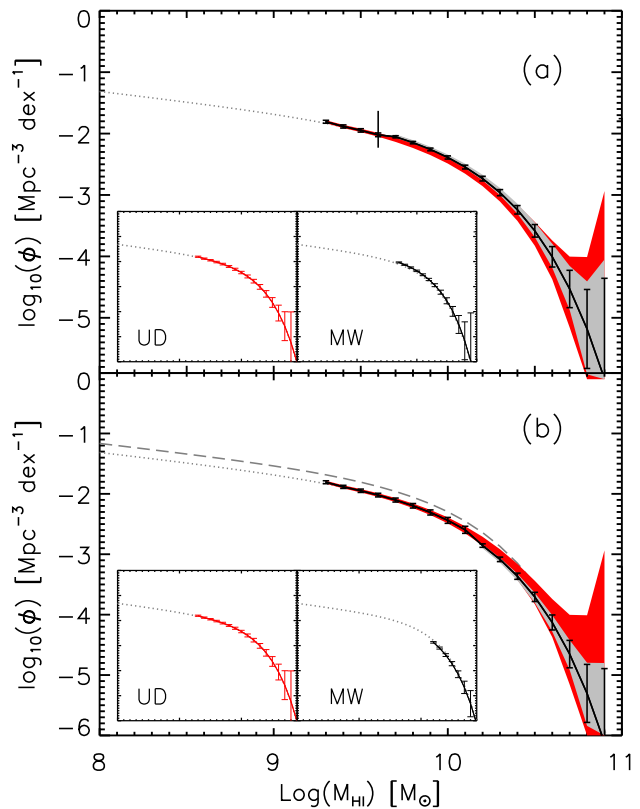}
\caption{The HIMF resulting from combining \HI\ data from the UD (red)
  and MW (grey) 
components of the Optimised (top panel) and Ideal surveys (bottom panel)
at $0.35<z<0.45$. The larger area of the MW component in Ideal
compared to the Optimised case improves the constraints at the
high-mass end. The short vertical line in the top panel marks the
limiting \HI\ mass if the sensitivity of the current Band 1 is reduced
by a factor of two, and the dashed line in the bottom panel indicates
the HIMF if allowed to evolve with redshift. The
dotted line is the $z=0$ HIMF as in previous figures. The x-axis in
the main panel and the insets span the same range.}
\label{fig:best1_best2_HIMF_z04_inset}
\end{figure}

At higher redshift, $0.35<z<0.45$, Fig.~\ref{fig:best1_best2_HIMF_z04_inset}
shows that the Optimised and either of the Ideal surveys both
incorporate the UD component, but lose significant contribution from
the W survey, which only detects galaxies in the final few \HI\ mass
bins. The short vertical line in the top panel indicates the limiting \HI\ mass reached
if the sensitivity of the current Band 1 receiver is reduced by a
factor of two. The bottom panel shows that the larger area of the Ideal
MW tier reduces the uncertainties at the high-mass end of the
HIMF. If we allow the normalization of the HIMF to evolve with
redshift such that by $z=1$ the 
integral is twice that at $z=0$, we get the dashed curve shown in the
bottom panel of Fig.~\ref{fig:best1_best2_HIMF_z04_inset}. The
  uncertainties in the HIMF determination for both cases 
are sufficiently small to distinguish between the evolving and non-evolving cases.

\section{Complementarity of \HI\ and continuum
  surveys}\label{sec:deepsurveys}

The benefits of observing commensally to minimize observational effort
while maximizing science return are clear. Provided sufficient advance
effort with respect to survey design and science requirements is
invested, it is possible to construct a coherent dataset from disparate surveys. 
With a major scientific effort such as the SKA, the accessibility of
previously unexplored parameter space is
expected to transform our understanding of a number of key areas of
astrophysics, so we must ensure we undertake surveys which will
provide the data needed to significantly advance our current
knowledge.

Here we explore the \textit{complementarity} of the data products resulting
from the commensal \HI\ and radio continuum observations, biases in
the resulting dataset, limitations
resulting from restricted sensitivity at different wavelengths, and
how information can be combined to circumvent these limitations.

\subsection{Low redshift \HI\ and Optical Data}\label{subsec:data}

With future observations, we will have simultaneous measures of both
the \HI\ reservoirs of galaxies, along with star formation rate
indicators from the radio continuum, over a cosmological redshift
range. Our current information, particularly \HI, is restricted to
$z\sim 0$. While limited, the existing data can be used to provide the context
within which we can expect the new observations to fit.

To understand how \HI, optical and radio continuum surveys will fit
together, we take advantage of the spatial overlap of
two individual, large-area surveys. Optical photometry and
spectroscopy are provided by the Sloan Digital Sky
Survey (SDSS, \citealt{York2000}). The \HI\ data are from the
Arecibo Legacy Fast ALFA survey (ALFALFA, \citealt{Giovanelli2005}) 
$\alpha$.40 source catalogue from \citet{Haynes2011}, which
covers $\sim$2800 deg$^2$ within the SDSS imaging footprint, and
contains 15\,041 secure 
extragalactic sources. ALFALFA is flux-limited to $\sim$0.47 Jy\ \kms,
while the spectral range allows observations of galaxies to
$z=0.06$. From this catalogue we extract the \HI\ masses, \MHI, from the \HI\
profiles, along with the galaxy redshifts provided within the ALFALFA
data release.

The ALFALFA team has crossmatched the $\alpha$.40 catalogue to the
SDSS Data Release 7 (DR7, \citealt{Abazajian2009}) by hand to identify
the most likely optical counterparts for the \HI\ detections. 11\,740 of the
ALFALFA galaxies have clear optical counterparts in the DR7 imaging,
providing optical photometry, galaxy sizes, and for the majority of
the objects, spectroscopic redshifts. We refer to this set of
galaxies as the ALFALFA--SDSS sample. 

At the intersection of ALFALFA and SDSS, we have, for a large
number of galaxies, information regarding the neutral gas reservoir,
the existing stellar properties, and a measure of the star formation
rate from optical photometry.
Even though ALFALFA is restricted to $z<0.06$, the galaxies
contained within the survey span a wide range of \HI\ masses, stellar
masses, and star formation rates, and provide a
useful low redshift frame of reference for extrapolating to the higher
redshifts and greater depths which will be probed by upcoming surveys.

\subsection{\HI\ and radio continuum selection of galaxies}\label{subsec:biases}

As described in detail in \citet{Huang2012}, \HI\ selection results in
a highly biased galaxy sample, preferentially including blue, star
forming objects. Constructing a galaxy sample flux-limited in the radio continuum
also results in a biased census, composed of two populations of
objects. The high luminosities are dominated by galaxies hosting active galactic nuclei
(AGN), whereas at lower luminosities, the star-forming galaxies become
more prevalent (for example, \citealt{Wilman2008}, \citealt{Padovani2009},
\citealt{Massardi2010}, \citealt{McAlpine2013}). Continuum
surveys with flux limits sensitive enough to probe below the 
AGN regime will have significant numbers of star-forming galaxies,
which are the same galaxies selected by deep \HI\ surveys. So although
both \HI\ and continuum selection result in biased galaxy samples,
they are biased toward the same galaxies. 

The galaxy mass function derived from the Galaxy And Mass Assembly
(GAMA, \citealt{Driver2011}) shows that at the highest stellar masses,
the galaxy counts are dominated by red galaxies \citep{Baldry2012}, which tend to be
\HI-poor and have little star formation activity. However, at stellar
masses \Mstellar$<10^{10.3}$\,\Msun, the galaxy counts are
dominated by blue galaxies, which is also where the bulk of star formation activity
occurs, and is exactly the population of objects both \HI\ and
continuum selection are biased toward. 

\subsection{Optical Imaging and Spectroscopy}\label{subsec:fluxlim}

A multi-wavelength aspect, combining several individual surveys at
different wavelengths, is becoming increasingly common with current
and future surveys, including upcoming radio-based observations.
These ancillary data, primarily optical imaging and spectroscopy, 
provide additional information for the galaxy population under
examination. However, the optical observations impose their own flux
limits on a given survey in addition to the \HI\ and continuum
sensitivities imposed by the radio wavelength observing
parameters. Imaging large areas of sky to significant depths in order
to obtain galaxy colours and morphological information is well within
the capabilities of current optical instrumentation, as proven by, for
example, the Canada-France-Hawaii Telescope Legacy Survey
(CFHTLS\footnote{\url{http://www.cfht.hawaii.edu/Science/CFHTLS/}}),
the Cosmic Evolution Survey (COSMOS, \citealt{Scoville2007}), 
UltraVISTA (\citealt{McCracken2012}), the VISTA Deep Extragalactic
Observations survey (VIDEO, \citealt{Jarvis2013}), and Dark Energy Survey
(DES\footnote{\url{http://www.darkenergysurvey.org}}) and do not
impose significant limitations on the galaxy sample to be observed. 
Upcoming surveys such as the Large Synoptic Survey Telescope (LSST,
\citealt{Ivezic2008}), Euclid (\citealt{Laureijs2012}), and
WFIRST\footnote{\url{http://wfirst.gsfc.nasa.gov}} actively
incorporate surveys at other wavelengths in their planning strategy.

Large area, flux-limited spectroscopic surveys, however, require
extensive telescope time and effort. While photometric redshifts
(photo-$z$), with accuracies of $\Delta z \sim 0.05$ are sufficient
for a number of scientific applications (the evolution of luminosity
functions, for example), the accuracy of 
spectroscopic redshifts ($\Delta z \sim 0.0003$) are required for
investigations of detailed galaxy environment and spectral stacking
(\citealt{Maddox2013}). The acquisition of sufficient 
spectroscopic redshifts to supplement 
upcoming \HI\ and continuum surveys is the topic of much discussion
and effort (see a good overview in \citealt{Meyer2015}), and here we
investigate the effect of imposing an optical flux limit on a galaxy
sample initially flux-limited in \HI\ as a function of redshift, and
what other information we can use to alleviate the effects. 

\subsubsection{Scaling Relations and Flux Limits}\label{subsec:scaling}

\begin{figure}
\includegraphics[width=\columnwidth]{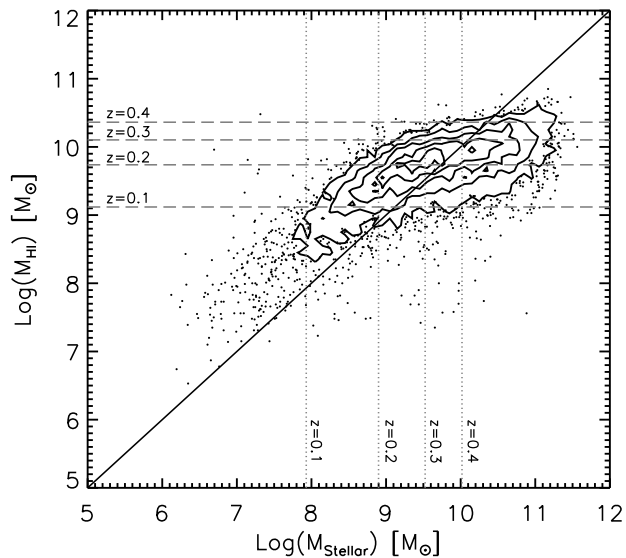}
\caption{\MHI\ as a function of \Mstellar\ for the ALFALFA--SDSS
  galaxies, in units of \Msun. The diagonal black solid line marks the
  1-to-1 relation. The vertical grey dotted lines indicate the stellar
  mass limits resulting from an optical flux limit of $r=22$ at various
  redshifts as described in Section~\ref{subsec:fluxlim}, while the
  horizonal grey dashed lines indicate the 
  \HI\ mass sensitivity of the CHILES survey at $z=0.1-0.4$, outlined
  in Section~\ref{sec:chiles}.}
\label{fig:A_MHI_Mstellar}
\end{figure}

Scaling relations illustrate general trends between physical
properties for an ensemble of galaxies. One such relation at $z\sim 0$ is shown
in Fig.~\ref{fig:A_MHI_Mstellar}, indicating how \HI\ mass, from the
ALFALFA catalogue, and stellar mass, from the MPA--JHU catalogue
(\citealt{Brinchmann2004}, \citealt{Tremonti2004}), 
are related for the \HI-selected ALFALFA--SDSS galaxies. At low stellar masses,
below \Mstellar$<10^9$\,\Msun, the galaxies are \HI-rich, lying above
the diagonal line marking the 1-to-1 relation. At high stellar masses,
the galaxies become increasingly \HI-poor. Extensive discussion
regarding the features of this relation can be found in \citet{Maddox2015}.
There are many galaxies not
detected by ALFALFA which would lie in the bottom right corner of
Fig.~\ref{fig:A_MHI_Mstellar}, at high stellar mass and low \HI\
mass. These are the absent red sequence galaxies not included in an
\HI-selected sample. While a small number of early-type galaxies
are found to have significant masses of \HI\ \citep{Serra2012}, 
the contribution from these galaxies to the \HI\ mass census is negligible.

Known scaling relations can also be useful for
converting observable quantities, such as brightness, into
quantities that are difficult to measure either directly or
indirectly. For example, a galaxy's stellar content can be estimated
using only it's brightness and colour by employing the known
correlation between the optical luminosity and colours of a galaxy and
its stellar mass (\citealt{Bell2003}). We use this relation to convert optical flux limits
to approximate limiting stellar masses, as a function of redshift, to
illustrate the effect these flux limits have on an \HI-selected galaxy
sample.

Due to the biased nature of the ALFALFA--SDSS galaxies, we derive
a Luminosity--\Mstellar\ relation specifically for this dataset. As
the \HI-selected galaxies are nearly all
blue cloud galaxies, we remove the colour dependence of the usual
Luminosity--\Mstellar\ relation by assuming all galaxies have the same
representative colour. This approximation is sufficient for the
illustration presented here. We find:

\begin{equation}
\mathrm{Log(M_{\ast})} = -0.546(M_r) - 1.051\\
\label{eq:L_Mstellar}
\end{equation}

\noindent where \Mstellar\ is in units of solar masses, and $M_r$ is
the absolute $r$-band magnitude. We then use this relation to convert
a given optical flux limit into an approximate redshift-dependent stellar mass
limit, incorporating a $k$-correction appropriate for blue
galaxies. The calculation is only an approximation, because it uses
the relation derived from the $z=0$ population and assumes no
evolution with redshift. 

Shown on Fig.~\ref{fig:A_MHI_Mstellar} are vertical lines indicating the 
effective stellar mass limits resulting from imposing a flux limit of
$r<22$, appropriate for an optical spectroscopic campaign, on the
$z=0$ \HI-selected galaxy ensemble. The distribution of galaxies in
this figure is to provide the low redshift reference for the
populated regions of parameter space, which will evolve with redshift
in an unknown way, as we currently cannot make such a figure at
$z>0$. The limits at increasing redshift then illustrate which parts of
the parameter space are no longer accessible.

At $z=0.1$, all galaxies with
\Mstellar$>10^8$\,\Msun\ are brighter than $r=22$, but galaxies 
are rapidly lost with increasing redshift, leaving only objects with
\Mstellar$>10^{10}$\,\Msun\ bright enough for spectroscopy at
$z=0.4$. Note that these limits ignore the complicating factor of
galaxy surface brightness, which works against observations of disk
galaxies, particularly when viewed face-on. Relaxing the optical flux
limit to $r=24$, more appropriate for optical photometry 
alone, shifts the effective stellar mass limits lower by an order of
magnitude. This fainter flux limit is relevant for obtaining
photometric redshifts with current facilities over moderately wide areas.

Also shown on Fig.~\ref{fig:A_MHI_Mstellar} are horizontal lines
denoting the 5-$\sigma$ \HI\ mass sensitivity of CHILES as a function
of redshift (see Section~\ref{sec:chiles} for details). The
important detail to note is that for a given \HI\ mass, for example
\MHI$=10^9$\,\Msun, the stellar masses range from
$10^8<$\Mstellar$<10^{11}$\,\Msun.  Even at $z=0.3$, galaxies of
\Mstellar$\sim 10^9$\,\Msun\ will be detected in the most \HI-rich
cases, a factor of more than three lower in stellar mass than is accessible 
to optical spectroscopy at the same redshift, unless the objects have
very strong emission lines. Thus, selection to a
given \HI\ flux limit is effective 
at identifying galaxies with a wide range of stellar masses, whereas
selection incorporating an optical flux limit is strongly biased
toward the most massive stellar systems, particularly at increasing
redshift. In fact, the \HI\ observations themselves are able to provide redshifts
for these galaxies below the stellar masses accessible to optical
spectroscopy.

\subsection{Continuum and Spectral Stacking}\label{subsec:stacking}

In order to explore the average galaxy population below
the detection threshold of a given survey, statistical co-adding, or
stacking, is often employed. Stacking in the radio continuum has been
successfully performed by a number of authors (for example, \citealt{White2007},
\citealt{Garn2009}, \citealt{Karim2011}, \citealt{Zwart2015}, or see \citealt{ZwartSKA2015}
for a good overview). The spatial coordinates of the galaxies are provided by deep
optical or near-infrared imaging, and redshifts are required to convert fluxes
into luminosities.

With the commensal \HI\ and continuum observations, the galaxy
population selected in \HI, rather than at optical wavelengths, can be
used as the basis for a star formation census over the redshift range
probed by the spectral observations. Not only does the \HI\
selection provide a galaxy sample less biased by stellar mass, but the
\HI\ detections themselves provide the redshifts of the
galaxies, without the need for additional expensive spectroscopy,
which quickly becomes unfeasible for large numbers of galaxies with \Mstellar\ $\sim
10^9$\,\Msun\ at $z>0.2$. 

\subsection{Case study: CHILES and CHILES Con Pol}\label{sec:chiles}

As described in Section \ref{subsec:biases}, observations in \HI\ and
radio continuum tend to select the same galaxy population, so within
a single observation, we gain information about both the neutral gas content and
star formation activity in the same set of galaxies. We investigate
a pair of complementary spectral line and radio continuum surveys 
currently underway with the VLA.

The COSMOS \HI\ Large Extragalactic Survey
(CHILES\footnote{\url{http://chiles.astro.columbia.edu/}})
project is a deep, 1000 hour observation of a single pointing.
The instantaneous frequency coverage of
1450--970 MHz allows \HI\ detections over the unprecedented redshift range
$0<z<0.45$, which will only be surpassed by SKA-era facilities. Two
semesters of observations are 
already completed, and the feasibility is demonstrated by the pilot project,
spanning $0<z<0.193$ \citep{Fernandez2013}. The same 1000 hours of
observations are being shared with the CHILES Continuum Polarization
(CHILES Con Pol\footnote{\url{http://www.chilesconpol.com}},
\citealt{Hales2014}) team, who are exploiting the flexibility and
power of the upgraded VLA correlator to simultaneously observe in
four 128 MHz continuum bands to create a deep continuum map of the
same area of sky.

The total observing time for CHILES was chosen in order to detect a galaxy with
\MHI$\sim 3\times 10^{10}$\,\Msun\ at the highest accessible
redshift. This corresponds to a 5-$\sigma$ flux limit of 0.0295 Jy \kms,
which translates directly to a redshift-dependent \HI\ mass limit via:

\begin{equation}
M_{\mathrm{HI}} = 2.356\times10^5\, D_{L}^2\,(1+z)^{-1} \int S_{v} \,d v \\
\label{eq:HImass}
\end{equation}

\noindent where \MHI\ is in solar masses, the luminosity distance to
the galaxy, $D_L$, is in Mpc, and the integral is the
total flux in Jy \kms. The $(1+z)$ factor accounts for the difference
between the observed and rest-frame profile width. Assuming an \HI\
profile width of 150 \kms, the minimum
detectable \HI\ mass at $z=$ 0.1, 0.2, 0.3 and 0.4 are marked as horizontal
dashed lines on Fig.~\ref{fig:A_MHI_Mstellar}. At $z=0.1$, galaxies
with \MHI$\sim 10^9$\,\Msun\ are easily detectable, thus including the
bulk of the \HI\ mass. The stellar masses for these galaxies spans
$10^8<$\Mstellar$<10^{11}$\,\Msun. Even at $z=0.2$, the \HI\ mass limit
is still below the break mass of the $z=0$ \HI\ mass function, at
$10^{9.96}$\,\Msun\ \citep{Martin2010}.

In reality, these mass limits have some variation for a given flux
limit, as the resulting detection limit is dependent on the HI profile width, as
determined by not only the mass of the galaxy, but it's inclination to
the line of sight. For a given flux limit, galaxies seen face-on
can be detected to lower mass limits than galaxies viewed edge-on, by
about 0.3 dex. Thus, a flux limit doesn't result in a hard lower mass limit, but
rather a lower mass limit with a decreasing number of galaxies to
lower masses seen increasingly face-on. The limiting \HI\ masses
marked on Fig.~\ref{fig:A_MHI_Mstellar} should then only be used for guidance.

In a similar manner to \HI, the expected flux limit of
0.6$\mu$Jy per beam in continuum for CHILES Con Pol can be converted to
an equivalent detectable star formation rate as a function of 
redshift. To convert the 1.4 GHz continuum flux, $f$, into luminosity,
we use:

\begin{equation}
L = \frac{f D_L^2 (1+z)^{\alpha -1}}{1\times 10^{26}} \\
\label{eq:radlum}
\end{equation}

\noindent where the luminosity, $L$, is in W Hz$^{-1}$ sr$^{-1}$ and assuming
a power-law spectral index $S_{\nu} \propto \nu^{-\alpha}$ and
$\alpha=0.7$. Then, using the relation to convert 1.4 GHz luminosity to a star
formation rate, we use the relation from \citet{Murphy2011}:

\begin{equation}
{\mathrm{SFR}} (\mathrm{M_{\odot}}\ yr^{-1}) = 4\pi\ 6.35\times 10^{-22} L \\
\label{eq:SFR_1}
\end{equation}

\noindent Even at $z=0.4$, CHILES Con Pol will be sensitive enough to detect
star formation rates of $\sim$ 1 \Msun\ year$^{-1}$. Thus, for the
first time, we will have information about moderate levels of star
formation, along with the \HI\ reservoirs, in galaxies over a range of
redshifts. Due to the computing capacities now available, this
information can be collected with one set of observations.

With the relatively small field of view of the VLA, $\sim$0.5 deg in
diameter, and restricted redshift range of $0<z<0.5$ for \HI, the
number of galaxies contained within the observed volume is manageably
small, on the order of 300 galaxies. The
CHILES pointing is within the COSMOS footprint, thus deep
multi-wavelength imaging (\citealt{Scoville2007},
\citealt{McCracken2012}), along with extensive spectroscopy, already 
exist (\citealt{Lilly2007}, \citealt{Lilly2009},
\citealt{Garilli2008}, \citealt{Coil2011}, \citealt{Cool2013},
\citealt{Ahn2014}). These catalogues have been conveniently
amalgamated, and for the zCOSMOS spectra, reprocessed, by
\citet{Davies2015}. This indicates the amount of ancillary data
required to support upcoming radio surveys. With the larger fields
of view and wider frequency coverage of SKA-era facilities, the ancillary data effort
becomes even more substantial.

\section{Summary}\label{sec:summary}

Despite \HI\ ultimately providing the raw material required for star
formation, little is known about how galaxies acquire and consume this
fuel over cosmological timescales. Conversely, the star formation
density in the Universe has been traced with a number of techniques to
the highest observable redshifts. With current and upcoming radio
facilities, enabled by advances in engineering and computing, we will,
for the first time, be able to trace both the \HI\ 
reservoir and the star formation activity in the same galaxies over a
cosmologically significant timescale, when the SFR is dropping rapidly
but the neutral gas content remains relatively stable.

Careful planning of commensal observations enables multiple
science cases to benefit from the same data, and maximises the
scientific return with no impact on telescope time. The MeerKAT
programmes LADUMA and MIGHTEE are an excellent example of survey tiers
well matched in terms of relative depth and area coverage, creating a
comprehensive \HI\ dataset. Conversely, the three \HI\ survey tiers
suggested by the SKA \HI\ science working group to be undertaken with
the SKA1-MID instrument require significant rearrangement in terms of
area and depth to create a coherent dataset. The definition of the
frequency bands is a critical aspect when designing the tiered survey,
especially for \HI\ science. We find that alternative frequency band
definitions, either the suggested alteration of Bands 1 and 2, or a
new single-octave band covering 1200--600MHz, require less extreme
survey tiers, and less observing time, in order to optimally cover the
redshift--\MHI\ parameter space. 

In addition to \HI\ and continuum observations being highly commensal,
the data products are also complementary. \HI\ selection results in a
galaxy sample biased toward blue, star-forming galaxies, which are the
same galaxies that dominate radio continuum selection at faint flux
limits. \HI\ selection also results in a sample less biased in terms
of stellar mass than the effective optical flux limit that results from
the additional requirement of obtaining spectroscopic redshifts. In
fact, data from the radio spectral line observations are able to
provide redshifts for gas-rich galaxies without needing additional
optical spectroscopy.

Commensal observing will become more common as new facilities with
expanded capabilities come online, and large programmes compete for
ever increasing time allocations. Maximising the science return for 
many separate projects while minimising observational effort will be
essential to achieve the ambitious goals set out by each team, whose
individual survey plans would not be realised otherwise. The current
investigation shows that with adjustments of survey parameters,
including area and depth, a coherent strategy satisfying several communities is
possible, provided advance effort is committed to the planning stages.








\section*{Acknowledgements}

The Arecibo Observatory is operated by SRI International under a
cooperative agreement with the National Science Foundation
(AST-1100968), and in alliance with Ana G. M{\'e}ndez-Universidad
Metropolitana, and the Universities Space Research Association. 
We acknowledge the work of the entire ALFALFA collaboration
team in observing, flagging, and extracting the catalogue of galaxies used in this
work. 

Funding for the SDSS and SDSS-II was provided by the Alfred P. Sloan
Foundation, the Participating Institutions, the National Science
Foundation, the U.S. Department of Energy, the National Aeronautics
and Space Administration, the Japanese Monbukagakusho, the Max Planck
Society, and the Higher Education Funding Council for England. The
SDSS was managed by the Astrophysical Research Consortium for the
Participating Institutions. The SDSS Web Site is
http://www.sdss.org/.

The SDSS is managed by the Astrophysical Research Consortium for the
Participating Institutions. The Participating Institutions are the
American Museum of Natural History, Astrophysical Institute Potsdam,
University of Basel, University of Cambridge, Case Western Reserve
University, University of Chicago, Drexel University, Fermilab, the
Institute for Advanced Study, the Japan Participation Group, Johns
Hopkins University, the Joint Institute for Nuclear Astrophysics, the
Kavli Institute for Particle Astrophysics and Cosmology, the Korean
Scientist Group, the Chinese Academy of Sciences (LAMOST), Los Alamos
National Laboratory, the Max-Planck-Institute for Astronomy (MPIA),
the Max-Planck-Institute for Astrophysics (MPA), New Mexico State
University, Ohio State University, University of Pittsburgh,
University of Portsmouth, Princeton University, the United States
Naval Observatory, and the University of Washington. This research has
made use of NASA's Astrophysics Data System. 

This work has benefitted from useful discussions, particularly with
Ian Heywood and Chris Hales. MJJ acknowledges support from SKA South
Africa and STFC. We thank the anonymous referee for helpful comments
which improved this paper. 









\bsp	
\label{lastpage}
\end{document}